\documentclass[conference]{IEEEtran}
\IEEEoverridecommandlockouts

\usepackage{cite}
\usepackage{amsmath,amssymb,amsfonts}
\usepackage{algorithmic}
\usepackage{graphicx}
\usepackage{textcomp}
\usepackage{xcolor}
\usepackage{booktabs}
\usepackage{multirow}
\usepackage{pifont}
\usepackage{tcolorbox}
\usepackage{subfigure}
\usepackage{colortbl}
\usepackage{makecell}
\def\BibTeX{{\rm B\kern-.05em{\sc i\kern-.025em b}\kern-.08em
    T\kern-.1667em\lower.7ex\hbox{E}\kern-.125emX}}
\begin{document}

\title{PerProb: Indirectly Evaluating Memorization in Large Language Models}

\author{
    \IEEEauthorblockN{
        Yihan Liao\IEEEauthorrefmark{1},
        Jacky Keung\IEEEauthorrefmark{1},
        Xiaoxue Ma\IEEEauthorrefmark{2},
        Jingyu Zhang\IEEEauthorrefmark{1}, and
        Yicheng Sun\IEEEauthorrefmark{1}
    }
    \IEEEauthorblockA{
        \IEEEauthorrefmark{1}Department of Computer Science, City University of Hong Kong, Hong Kong, China\\
        \IEEEauthorrefmark{2}Electronic Engineering and Computer Science, Hong Kong Metropolitan University, Hong Kong, China\\
        \{yihanliao4-c, jzhang2297-c, yicsun2-c\}@my.cityu.edu.hk, jacky.keung@cityu.edu.hk, kxma@hkmu.edu.hk
    }
}

\maketitle

\renewcommand{\thefootnote}{\fnsymbol{footnote}}
\footnotetext[2]{Xiaoxue Ma is the corresponding author.}
\renewcommand{\thefootnote}{\arabic{footnote}}

\begin{abstract}
The rapid advancement of Large Language Models (LLMs) has been driven by extensive datasets that may contain sensitive information, raising serious privacy concerns. One notable threat is the Membership Inference Attack (MIA), where adversaries infer whether a specific sample was used in model training. However, the true impact of MIA on LLMs remains unclear due to inconsistent findings and the lack of standardized evaluation methods, further complicated by the undisclosed nature of many LLM training sets. To address these limitations, we propose \textit{PerProb}, a unified, label-free framework for indirectly assessing LLM memorization vulnerabilities. \textit{PerProb} evaluates changes in perplexity and average log probability between data generated by victim and adversary models, enabling an indirect estimation of training-induced memory. Compared with prior MIA methods that rely on member/non-member labels or internal access, \textit{PerProb} is independent of model and task, and applicable in both black-box and white-box settings. Through a systematic classification of MIA into four attack patterns, we evaluate \textit{PerProb}'s effectiveness across five datasets, revealing varying memory behaviors and privacy risks among LLMs. Additionally, we assess mitigation strategies, including knowledge distillation, early stopping, and differential privacy, demonstrating their effectiveness in reducing data leakage. Our findings offer a practical and generalizable framework for evaluating and improving LLM privacy.
\end{abstract}


\section{Introduction}
Large Language Models (LLMs) play a critical role in Natural Language Processing (NLP) with wide application \cite{min2023recent, alshahwan2024automated, bairi2024codeplan}. However, the widespread adoption of LLMs in various applications has raised significant privacy concerns, as these models are trained on extensive datasets that may include sensitive information \cite{meeus2024did}. Membership Inference Attack (MIA) is one of the most prominent privacy risks, where an adversary attempts to determine whether specific data is part of a model's training set \cite{shokri2017membership}. Such data leakage can lead to severe consequences, including personal privacy violations and intellectual property breaches \cite{hu2022membership}. This concern is further reinforced by the General Data Protection Regulation (GDPR) in the European Union, which considers the inference of training data membership as a privacy violation, even if the data itself is not directly disclosed \cite{voigt2017eu}. In practice, successful MIA attacks may expose private health records, personal conversations, or proprietary code snippets, raising serious ethical and legal concerns in both commercial and governmental artificial intelligence deployments.

However, the effectiveness of MIA on LLMs remains a controversial topic. Some studies have demonstrated successful attacks \cite{carlini2021extracting, mattern2023membership, shi2023detecting, meeus2024did, puerto2024scaling}, while others argue that MIAs perform only marginally better than random guessing in most settings \cite{duan2024membership, das2024blind, maini2024llm}. This inconsistency stems from differing evaluation methodologies and the challenge of obtaining ground-truth member/non-member labels, as many widely-used LLMs (e.g., GPT series \cite{radford2019language}) do not publicly disclose pretraining data.

To address this challenge, we propose \textit{PerProb}, a unified framework that introduces \textbf{Per}plexity ($PPL$) and average log \textbf{Prob}ability ($\lambda(W)$) as two metrics to indirectly assess LLM memorization behaviors. \textit{PerProb} compares model outputs generated under identical prompts before and after training, using shifts in $PPL$ \cite{carlini2021extracting} and $\lambda(W)$ \cite{mitchell2023detectgpt} to reveal training-induced memory effects. Unlike prior MIA methods, \textit{PerProb} does not rely on pre-defined labels or internal access, and instead infers memorization based solely on output distributions. In addition to generation tasks, we apply \textit{PerProb} to classification tasks by constructing member and non-member sets with matched distributions, allowing direct evaluation of MIA effectiveness across different settings.

To systematically evaluate LLM privacy, we divide MIA into four attack patterns, reflecting different black-box and white-box threat models \cite{sablayrolles2019white}. Black-box attacks rely only on model inputs and outputs, while white-box attacks leverage internal information such as architecture or training data access \cite{liu2023gradient, wu2023understanding}. Our empirical results demonstrate that \textit{PerProb} can effectively uncover memorization behaviors in LLMs, particularly in smaller-scale models. On classification tasks, we observe average F1-scores around 70\% across the four attack patterns. Both black-box and white-box attacks reveal substantial privacy risks, highlighting the importance of systematic defense evaluations. Accordingly, we further assess three mitigation strategies: Knowledge Distillation (KD) \cite{gou2021knowledge}, Early Stopping (ES) \cite{dodge2020fine}, and Differential Privacy (DP) \cite{abadi2016deep}, all of which show varying levels of success in mitigating MIA.

\textbf{Contributions.} The main contributions are as follows:
\begin{enumerate}
    \item We propose \textit{PerProb}, a label-free framework for indirectly evaluating LLM memorization based on perplexity and average log probability in generation tasks.

    \item We provide a unified assessment across generation setting under four MIA attack patterns, demonstrating \textit{PerProb}’s effectiveness across models and datasets.
    
    \item We evaluate mitigation strategies including KD, ES, and DP, and show their capability to reduce privacy leakage in practical LLM deployments.
\end{enumerate}

\section{Background}

Membership Inference Attack (MIA) refers to the ability to determine whether a particular data sample was included in the training set of a target model, thereby posing serious privacy threats \cite{shokri2017membership}. Early work by Shokri et al. proposed the foundational shadow model framework, demonstrating MIA feasibility on various machine learning models. Subsequent efforts categorized MIA into \textit{black-box} and \textit{white-box} paradigms \cite{sablayrolles2019white}. In black-box attacks, adversaries can only query model inputs and observe outputs \cite{liu2023gradient}, while white-box attacks assume access to model parameters or gradients \cite{wu2023understanding}. Both paradigms have been extensively explored in conventional ML settings using attack signals such as confidence scores \cite{shokri2017membership}, prediction loss \cite{liu2022ml}, gradients \cite{liu2023gradient}, or intermediate representations.

With the advancement of deep learning, recent attention has shifted toward evaluating MIA risks on Large Language Models (LLMs). While LLMs have achieved impressive performance across NLP tasks, they are typically trained on massive, undisclosed corpora, raising concern that sensitive information may be memorized and exposed via inference. Carlini et al. first demonstrated the feasibility of data extraction from GPT-2 \cite{carlini2021extracting}, followed by other works such as Mattern et al. \cite{mattern2023membership} who proposed neighborhood-based attacks, and Shi et al. \cite{shi2023detecting} who attempted membership detection via prompt calibration. However, recent studies have raised skepticism about the consistency and significance of MIA against LLMs, arguing that many prior results are fragile and may rely on overfitted or task-specific setups \cite{duan2024membership, das2024blind, maini2024llm}.

Two core challenges make MIA against LLMs particularly difficult: \textbf{Challenge 1: Lack of membership labels.} Most LLMs, such as the GPT series, do not disclose their pre-training data, making it infeasible to obtain ground-truth labels for evaluating membership. Existing work circumvents this by heuristically selecting non-member data created after the model’s training cutoff date \cite{duan2024membership, meeus2024did}. However, such heuristics are unreliable, as post-cutoff data may be included via backchannels, and pre-cutoff data may be excluded. An alternative approach is to split benchmark datasets (e.g., AGNews) based on data distribution and assume non-overlap with the pretraining set \cite{mattern2023membership, fu2023practical}, but this assumption remains unverifiable. \textbf{Challenge 2: Generalization vs. memorization in LLMs.} LLMs are trained to generalize over vast corpora, making individual memorized examples harder to distinguish. While smaller models (e.g., GPT-2) may exhibit overfitting, modern LLMs tend to produce high-entropy predictions even for seen data, reducing the discriminative power of conventional MIA features such as confidence or loss \cite{duan2024membership, chen2024statistical}.

To address these challenges, some researchers have turned to open-source LLMs like Pythia, which disclose their training sets, allowing direct membership evaluation \cite{puerto2024scaling, ren2025self}. However, such models are not representative of widely deployed proprietary LLMs (e.g., ChatGPT), limiting the generalizability of these results. Furthermore, conventional MIA techniques do not extend well to settings without label supervision, underscoring the need for indirect, label-free MIA.

\textbf{Our Motivation.} To bridge this gap, we propose \textit{PerProb}, a unified, label-free framework for assessing LLM memorization under both black-box and white-box assumptions. Rather than relying on ground-truth membership labels, \textit{PerProb} compares perplexity ($PPL$) and average log-probability ($\lambda(W)$) of generated outputs across shadow and victim models. This indirect approach enables memory assessment across both generation and classification tasks without requiring access to training data labels. Table~\ref{tab:bg} summarizes representative MIA approaches, their assumptions, and limitations.

\begin{table*}[htbp]
\centering
\caption{Representative MIA Methods: Assumptions, Access, and Limitations}
\label{tab:bg}
\renewcommand{\arraystretch}{1.4}
\begin{tabular}{m{2.8cm}m{3cm}m{4.5cm}m{5cm}}
\toprule
\multicolumn{1}{c}{\textbf{Method Type}} & 
\multicolumn{1}{c}{\textbf{Assumptions}} & 
\multicolumn{1}{c}{\textbf{Required Access}} & 
\multicolumn{1}{c}{\textbf{Limitations in LLMs}} \\
\hline
\rowcolor{gray!25}
Confidence-based \cite{shokri2017membership} & Requires labeled member /non-member data & Model output probabilities (black-box) & Confidence often indistinguishable due to LLM generalization \\
\rowcolor{white}
Loss-based \cite{mattern2023membership} & Assumes lower loss on member samples & Loss scores or logits & Loss values are unstable and may not differ significantly in LLMs \\
\rowcolor{gray!25}
Gradient-based \cite{liu2023gradient} & Access to model gradients & Model parameters and gradients (white-box) & Impractical for large-scale, API-access LLMs; expensive computation \\
\rowcolor{white}
Shadow model \cite{shokri2017membership} & Train shadow models on similar data & Data generator and model architecture & Requires labeled data; inaccurate in real-world LLMs due to data mismatch \\
\rowcolor{gray!25}
Prompt-based \cite{shi2023detecting, fu2023practical} & Use calibrated prompts to infer membership & Model responses to crafted prompts & Performance highly dependent on prompt design and task \\
\rowcolor{white}
\textbf{PerProb (Ours)} & No label or internal access required & Only model outputs; applicable to both generation and classification & Indirect inference, assumes model behavior differs with training-induced memory \\
\bottomrule
\end{tabular}
\end{table*}

\section{Methodology}
\subsection{Models} 
To evaluate the memory ability of LLM, we focus on GPT-2 and GPT-Neo (1.3B and 2.7B), two representative models frequently studied in MIA research \cite{shi2023detecting, fu2023practical, ren2025self, mozaffari2024semantic, duan2024membership}. These models are selected for their non-transparency or partial transparency nature, demonstrating the versatility of the proposed \textit{PerProb} method. GPT-2, with an undisclosed pre-training set, represents a scenario where no ground truth member/non-member labels are available, requiring indirect evaluation \cite{kalyan2023survey}. GPT-Neo is primarily trained on The Pile \cite{gao2020pile}, which is a publicly available dataset. GPT-Neo introduces an intermediate transparency scenario where partial training data is available, but additional undisclosed data also exists. By leveraging \textit{PerProb}, which evaluates memory vulnerabilities through $PPL$ and $\lambda(W)$, we systematically analyze the differences in memorization responses under different LLMs. Findings provide generalized insights into the vulnerability of GPT-family LLMs and offer a reference framework for evaluating other non-transparent LLMs.

\subsection{Generation Task} 
\textit{PerProb} is a method designed to evaluate the memorization ability of LLMs by comparing the victim model’s responses to generated data before and after training. Unlike traditional MIA methods, \textit{PerProb} provides an indirect, label-free approach that relies on behavioral differences induced by training, making it especially suitable for models with limited training data transparency.

Specifically, \textit{PerProb} evaluates differences in two metrics: perplexity ($PPL$) and average log probability ($\lambda(W)$). $PPL$ captures the model’s uncertainty when predicting the next word, while $\lambda(W)$ reflects its overall confidence in generating a sequence. These metrics are computed as follows:
\begin{gather}
    PPL(W) = \exp\left( -\frac{1}{N} \sum_{i=1}^{N} \log p_{\theta}(w_i | w_{<i}) \right), \\
    \lambda(W) = \frac{1}{N} \sum_{i=1}^{N} \log p_{\theta}(w_i | w_{<i}),
\end{gather}
where $W$ denotes a token sequence of length $N$. Both metrics are derived from the model's token-wise probability distributions. The rationale behind \textit{PerProb} is grounded in information-theoretic intuition and prior MIA theory \cite{yeom2018privacy}. If a model has memorized specific training data, it tends to assign higher likelihoods to that data, resulting in lower perplexity and higher average log-probability. Let $x$ be a sample drawn from the training set. Then the expected loss on $x$ is lower than on unseen samples, as the model has directly optimized on it. This implies:
\begin{gather}
    PPL(x_{\text{member}}) < PPL(x_{\text{non-member}}), \\
    \quad \lambda(x_{\text{member}}) > \lambda(x_{\text{non-member}})
\end{gather}
In the absence of labeled member/non-member pairs, \textit{PerProb} leverages this statistical discrepancy between shadow and target models. By comparing the $PPL$ and $\lambda(W)$ values for data generated with identical prompts before and after training, we estimate whether the model’s output behavior reflects training-induced memory.

This design enables \textit{PerProb} to generalize the generation task, architectures, and access levels (black-box, white-box), offering a systematic way to assess memorization risk without requiring internal access or ground truth labels.

\subsection{Classification Task} 
For classification tasks, ground truth labels provide a direct approach to evaluate MIA. LLMs are used to predict class probabilities for input data, where the classification function is denoted as ${F}({X})={y}$, with ${X}$ as the input data and ${y}$ as the predicted probability vector. We adopt Random Forest (RF) and Multi-Layer Perceptron (MLP) as attack models for MIA evaluation due to their complementary strengths: RF offers robust parallel processing and handling of large-scale data \cite{sheykhmousa2020support}, while MLP is well-suited for learning complex, non-linear relationships \cite{taud2018multilayer}, both of which are suitable as the attack model against LLMs. To assess the effectiveness of attack models, we utilize three metrics, including precision, recall, and F1-score \cite{carlini2021extracting, liu2022ml}: $\text{Precision} = \frac{TP}{TP + FP}$, $\text{Recall} = \frac{TP}{TP + FN}$, $\text{F1-score} = 2 \cdot \frac{\text{Precision} \cdot \text{Recall}}{\text{Precision} + \text{Recall}}$. 
$TP$ is the number of member data correctly predicted as members, while $FP$ indicates the number of non-member data predicted as members.
$TN$ refers to the number of non-member data that are correctly predicted as non-members. $FN$ represents the number of member data that are incorrectly classified as non-members. 
By combining robust attack models and well-defined metrics, we provide a comprehensive evaluation framework for MIA on classification tasks.

\section{Threat Model Taxonomy} \label{threatModel}
\begin{figure*}[htbp]
    \centering
	\includegraphics[width=0.7\linewidth]{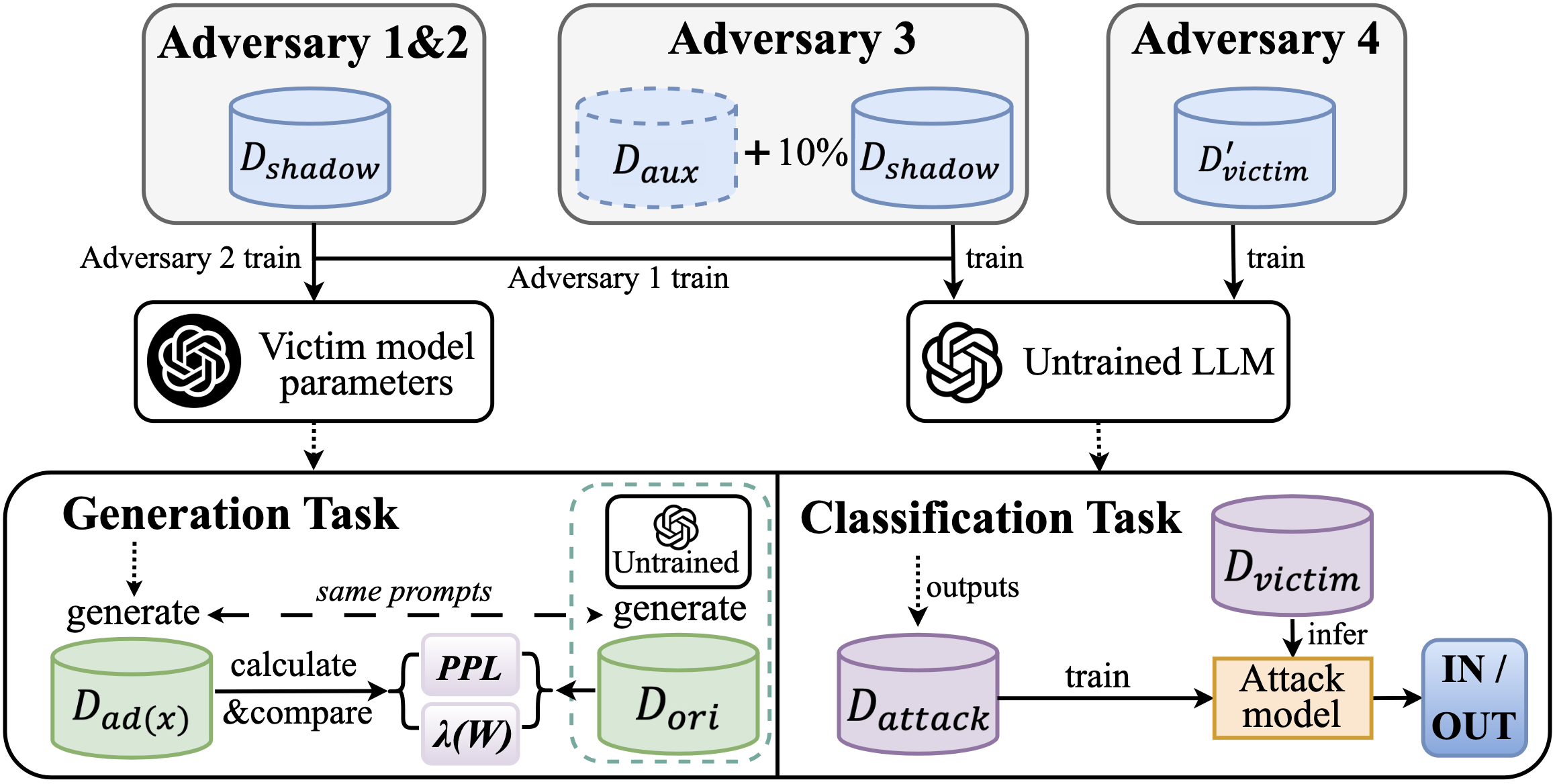}
\caption{Structure of four attack patterns on generation and classification tasks, respectively. \textbf{Adversary 1} and \textbf{Adversary 2} both use the same shadow datasets for training the $\mathcal{S}$, but Adversary 2 uses the $\mathcal{V}$ parameter. \textbf{Adversary 3} leverages the combination of the auxiliary dataset and 10\% shadow dataset to train the $\mathcal{S}$. \textbf{Adversary 4} utilizes partial victim datasets for training $\mathcal{S}$.}
\label{fig:overview}
\end{figure*}

A critical component in MIA is the shadow model ($\mathcal{S}$), which serves as a proxy for the victim model ($\mathcal{V}$), enabling adversaries to simulate the behavior of the $\mathcal{V}$ including approximating its decision boundaries and inferring membership, without direct access to it \cite{shokri2017membership, liu2022ml, hu2022membership}. We categorize MIA into four distinct patterns based on the characteristics of black-box and white-box attack frameworks and reconstruct the attacks across various scenarios. In each mode, we assume the adversary has different external knowledge about the training set or $\mathcal{V}$ to facilitate the inference process \cite{liu2022ml}. Fig.\ref{fig:overview} displays the disparities among four attack patterns, each attack being executed independently. 

\textbf{Adversary 1} represents a classic black-box attack, with an untrained LLM as the $\mathcal{S}$ that trains on shadow dataset ($D_{shadow}$), which mirrors the same distribution with the victim dataset ($D_{victim}$) to emulate the $\mathcal{V}$'s behavior. For the generation task, the untrained LLM and the $\mathcal{S}$ generate data respectively, which are preprocessed to remove noise, and \textit{PerProb} is applied in both generated datasets to compare and analyze the training-induced memorization. In the classification task, the outputs of $\mathcal{S}$ and $\mathcal{V}$ that separately train on $D_{shadow}$ and $D_{victim}$ are labeled with membership labels (training set as members, testing set as non-members), which integrated as $D_{attack}$ are used to train the attack model $\mathcal{A}$, which predicts membership for $D_{victim}$. This black-box setup is realistic and presents significant challenges for the adversary.

\textbf{Adversary 2} represents a white-box attack where the $\mathcal{S}$ shares parameters with the $\mathcal{V}$. In the generation task, $D_{shadow}$ is trained on the $\mathcal{S}$, which retains the $\mathcal{V}$’s parameters to generate data. \textit{PerProb} is then used to evaluate how shared parameters influence memorization. In the classification task, the $\mathcal{S}$ also utilizes the model parameters of $\mathcal{V}$ and trained on $D_{shadow}$ (split into member and non-member data), the outputs are used to train $\mathcal{A}$. This attack assumes full compromise of the $\mathcal{V}$’s parameters, making it the most severe but most challenging to achieve.

\textbf{Adversary 3} combines an auxiliary dataset ($D_{aux}$) with a small portion (10\%) of $D_{shadow}$ for training $\mathcal{S}$. The mixed dataset differs in distribution from the $\mathcal{V}$’s training data, enabling evaluation under limited overlap. In the generation task, the $\mathcal{S}$ trains on the mixed dataset and generates data for \textit{PerProb} computation, capturing how limited overlap affects memorization. In the classification task, the $\mathcal{S}$ trained on the mixed dataset provides outputs for training $\mathcal{A}$. This approach reduces reliance on $D_{shadow}$, offering a practical and scalable black-box attack.

\textbf{Adversary 4} represents a white-box attack where the adversary has access to a subset of $D_{victim}$ for training $\mathcal{S}$. Specifically, the $\mathcal{S}$ is trained directly on this partial data and generates data for the generation task, where \textit{PerProb} is computed. For classification tasks, the outputs of the $\mathcal{S}$ are the training inputs of $\mathcal{A}$. Despite potential privacy-preserving measures during pre-training, the use of public datasets and partial access to $D_{victim}$ make this attack feasible, providing adversaries with a strategy to extract extensive information.

\section{Experimental Setup}
\subsection{Datasets}
We use five publicly available text datasets to evaluate the memorization ability of LLMs and vulnerability to MIA: IMDB \cite{maas2011learning}, Agnews \cite{zhang2015character}, 20Newsgroup \cite{lang1995newsweeder}, Bank77 \cite{casanueva2020efficient}, and Web Of Science \cite{kowsari2017hdltex}. These datasets are selected to represent different types of text data, offering insights into how LLMs memorize information from various sources.

\begin{itemize}
    \item IMDB: contains 50,000 polarized movie reviews, equally balanced between positive and negative comments.
    \item Agnews: is a collection of news articles from over 2,000 sources, divided into 4 categories.
    \item 20Newsgroup: including 18,846 documents from 20 different newsgroups, distributed uniformly.
    \item Bank77: consists of 13,083 online banking queries focused on identifying client intent, with an average data length of 56.
    \item Web of Science (WOS): includes 46,985 documents from academic research papers, with 134 subcategories.
\end{itemize}

\subsection{Generation Task} 
In the generation task, we focus on three datasets: IMDB, Agnews, and WOS. These datasets are chosen for their diverse text types, which allows them to serve as $D_{aux}$ for one another in Adversary 3. This diversity enables us to evaluate the memorization patterns of LLMs across varying text domains.

For each dataset, we design tailored prompts to generate 1,000 data from original LLMs, denoted as $D_{ori}$. In Adversary 1, $D_{shadow}$ trained the $\mathcal{S}$ and generated 1000 text data, represented as $D_{ad1}$. In Adversary 2, the $\mathcal{S}$ integrates with the $\mathcal{V}$'s parameters and generates the dataset $D_{ad2}$. For Adversary 3, we use $D_{aux}$, alongside 10\% of $D_{shadow}$ to train the $\mathcal{S}$ and generate text data, denoted as $D_{ad3}$. Specifically, we pair IMDB with Agnews, IMDB with WOS, and Agnews with WOS, allowing evaluate the model’s ability to memorize data from different sources. In Adversary 4, we combine 10\% of $D_{victim}$ into $D_{shadow}$ for a white-box attack scenario. To analyze the memorization capacity of LLMs, we use \textit{PerProb} to measure changes between the untrained LLMs generated datasets ($D_{ori}$) and those generated datasets under each adversary configuration. 

\subsection{Classification Task} 
In the classification task, we use all five datasets to train $\mathcal{A}$. The attack model operates as a binary classifier, identifying whether a given data point belongs to the training set of the $D_{victim}$. The model we focus on in classification tasks in GPT-2, allows exploring MIA vulnerabilities in a controlled and interpretable setting, as its smaller size and simpler architecture compared to more advanced LLMs make it easier to analyze explicit memorization patterns. Additionally, as an earlier version of LLMs, GPT-2 may exhibit more substantial memorization effects due to less sophisticated training techniques, providing a valuable baseline for understanding MIA risks in LLMs. In this attack, we calculate the posterior probabilities of the data points based on the $\mathcal{S}$ and use these probabilities to train $\mathcal{A}$ to conduct MIA. If the F1-score of $\mathcal{A}$ surpasses 50\% (random guessing rate) \cite{shokri2017membership}, the attack is considered successful. In Adversary 3, we ensure the $D_{aux}$ has a similar number of categories to $D_{victim}$ to optimize the performance of the attack. For datasets with 2 or 4 categories, we randomly extract subsets of data from 20Newsgroup as the $D_{aux}$. For 20Newsgroup and Bank77, we select corresponding categories from the WOS dataset to ensure alignment in category numbers. Given the large number of categories in the WOS dataset, our analysis is confined to the first four datasets. In Adversary 4, the $\mathcal{S}$ obtains training data from $D_{victim}$, and the threshold of data acquisition spans from 0.1 to 0.5, simulating varying degrees of adversarial access to $D_{victim}$.

\subsection{Model Parameter}
For generation tasks, we use the four distinct adversaries $\mathcal{S}$ to reflect the memory capability of LLMs, where the configuration of these models has been discussed in Section~\ref{threatModel}. Each $\mathcal{S}$ is trained for 10 epochs with a learning rate set to 1e-6 to mitigate the risk of gradient explosion, ensuring stable training for the large-scale models. For classification tasks, we adapt the configuration of RF and MLP to the characteristics of the datasets. For datasets with fewer categories, such as IMDB, Agnews, and 20Newsgroup, the \textit{number of estimators} for RF is set to 100, balancing computational efficiency and model performance. For datasets with more complex structures or larger numbers of categories, like Bank77 and WOS, the \textit{number of estimators} is increased to 200 to capture the additional complexity. For MLP, the model architecture is tailored to the dataset’s complexity. In IMDB, Agnews, and 20Newsgroup, which have relatively fewer categories, the MLP is configured with three hidden layers to capture non-linear patterns without overfitting. In contrast, for Bank77 and WOS, which are characterized by a higher number of categories or more diverse content, the MLP is expanded to four hidden layers to enhance its capacity to learn more intricate relationships. 

\section{Evaluation}
\subsection{Generation Task}
\begin{figure*}[htbp]
    \centering
	\includegraphics[width=1\linewidth]{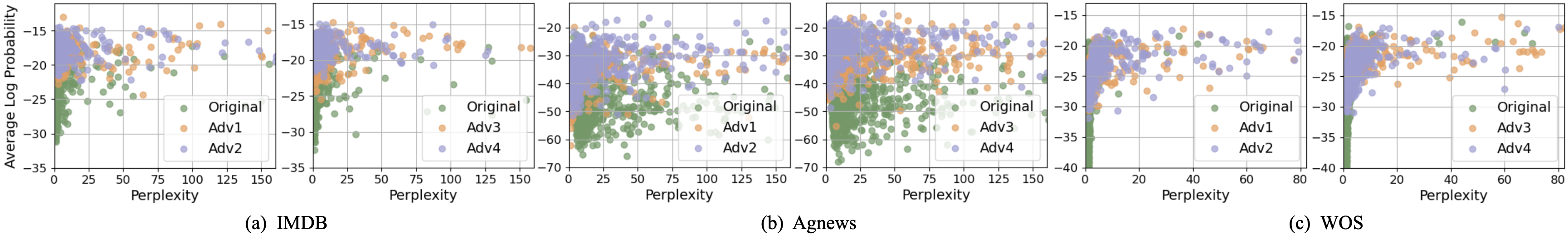}
\caption{The features of generated data on four attack patterns in GPT-2.}
\label{fig:gpt2}
\end{figure*}

\begin{figure*}[htbp]
    \centering
	\includegraphics[width=1\linewidth]{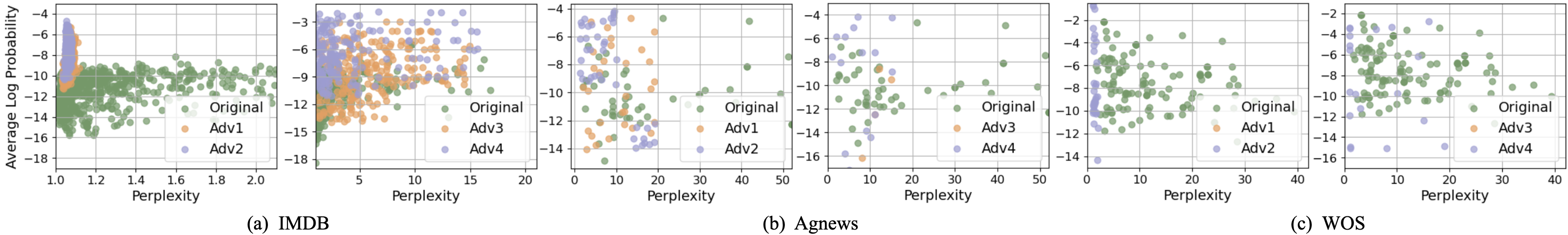}
\caption{The features of generated data on four attack patterns in GPT-Neo (1.3B).}
\label{fig:gpt1.3}
\end{figure*}

\begin{figure*}[htbp]
    \centering
	\includegraphics[width=1\linewidth]{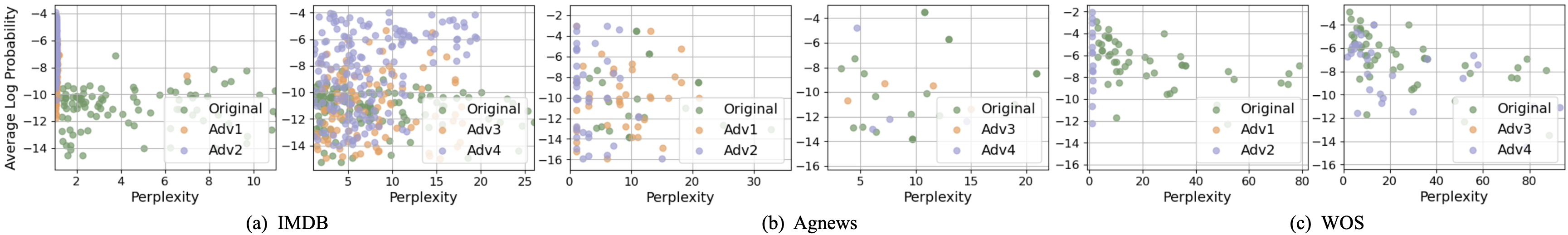}
\caption{The features of generated data on four attack patterns in GPT-Neo (2.7B).}
\label{fig:gpt2.7}
\end{figure*}

The experimental results, as illustrated in Figure \ref{fig:gpt2}, \ref{fig:gpt1.3}, and \ref{fig:gpt2.7}, offer compelling evidence of the susceptibility of LLMs to MIA across three different models (GPT-2, GPT-Neo 1.3B, and GPT-Neo 2.7B) under four attack patterns. The analysis focuses on \textit{PerProb} of the generated data on the $\mathcal{V}$ as key indicators to reflect LLMs' memorization. 

For attack patterns, Adversary 1 and Adversary 2, which are classic black-box and white-box attacks, generally show higher $\lambda(W)$ and lower $PPL$ compared to Adversary 3 and Adversary 4, as these attack patterns are closer to the $\mathcal{V}$’s distribution (Adversary 1 uses $\mathcal{S}$ trained on similar data distributions, while Adversary 2 uses $\mathcal{S}$ with $\mathcal{V}$ parameters). Conversely, Adversary 3 and Adversary 4, which involve $D_{aux}$ and partial victim data, exhibit higher $PPL$ and more instances of $-\infty$ $\lambda(W)$. The effectiveness of Adversary 3 in GPT-2 heavily depends on the theme between the $D_{aux}$ and the $D_{victim}$. $D_{aux}$ with similar semantics, such as 20Newsgroup and Agnews, enable $\mathcal{S}$ to mimic the $\mathcal{V}$ better, resulting in lower $PPL$ and higher $\lambda(W)$ on generated data.  While $D_{aux}$ with less alignment, like WOS, reduces $\mathcal{S}$ effectiveness due to distributional mismatches. However, GPT-Neo 1.3B shows a better performance on IMDB than Agnews in Adversary 3, leveraging its stronger generalization ability to generate high-confidence data even with less aligned auxiliary data, and tends to create more general data, which weakens the influence of theme consistency on the attack effect. 

When analyzing the three LLMs, we observe a trend that larger models produce more generated data with $\lambda(W) = -\infty$ and $PPL = \infty$, compared to smaller models like GPT-2 and GPT-Neo 1.3B. The more sparse distribution of scatters under GPT-Neo 1.3B and 2.7B suggests that even after training, the $\mathcal{V}$ has not memorized the training data significantly but retains strong generalization capabilities, failing to make high-confidence predictions on adversarially generated samples. 

This finding suggests that larger models may inherently resist memorizing specific training samples. While GPT-Neo generates more coherent and logical data due to its superior capacity, the observed trend of extreme values implies that its more substantial generative power does not necessarily lead to higher memorization risks. Instead, the model’s generalization capabilities allow it to focus on broader patterns rather than specific data points, enhancing its resilience against MIA.

Therefore, while smaller models’ lower $PPL$ and higher $\lambda(W)$ may initially appear to indicate higher vulnerability, the trend in larger models highlights an essential balance between generative power and privacy-preserving properties. This observation aligns with the theoretical understanding that larger models, despite their complexity, may generalize better across data and reduce overfitting, making it harder for adversaries to exploit direct memorization.

In addition to model-specific trends, dataset characteristics also play a critical role in determining MIA effectiveness. For instance, in the IMDB and Agnews datasets, the generated data under all attack patterns exhibits relatively consistent $PPL$ and $\lambda(W)$ across models. These datasets have well-defined semantic structures, as IMDB consists of binary sentiment labels, while Agnews categorizes news into four classes. Such uniformity likely helps both the $\mathcal{V}$ and $\mathcal{S}$ capture key features effectively, resulting in a more minor variance in attack outcomes. Conversely, datasets like WOS show more pronounced variations. With over 100 subcategories, WOS challenges both the $\mathcal{V}$ and $\mathcal{S}$ due to its high complexity. The $\mathcal{S}$ struggle to effectively replicate the $\mathcal{V}$’s behavior, leading to poorer attack performance and higher $PPL$. 

Besides, the performance of Adversary 3 highlights the role of dataset similarity. For example, in Agnews, the $D_{aux}$ (20 Newsgroup) shares thematic consistency with the $D_{victim}$, leading to lower $PPL$ and higher $\lambda(W)$ than datasets with less overlap (i.e., WOS and IMDB). This observation reinforces the importance of dataset alignment in MIA that $D_{aux}$ with structural or thematic similarity to $D_{victim}$ can significantly enhance attack effectiveness, while those with mismatched characteristics diminish attack success.

In summary, the results emphasize the interplay between attack pattern, model size, and dataset characteristics in determining MIA susceptibility. More extensive models like GPT-Neo 2.7B show greater resistance to MIA due to their generalization capabilities, producing more extreme values in metrics, which reflect reduced overfitting to specific training samples. Simultaneously, dataset complexity and thematic alignment between $D_{victim}$ and $D_{aux}$ significantly influence attack outcomes, with more straightforward or aligned datasets yielding stronger attacks. These findings collectively suggest that while LLMs’ memorization tendencies vary, both model architecture and dataset design are critical factors in mitigating MIA risks.

\begin{tcolorbox}[colback=gray!30, colframe=gray!70]
    \textbf{\textit{Finding 1}}: The effectiveness of MIA depends on training datasets and the generative capabilities of LLMs. Larger models exhibit stronger generalization and resistance to overfitting, while smaller models are more sensitive to MIA.
\end{tcolorbox}

\begin{figure}[htbp]
\centering
	\begin{subfigure}[IMDB]{
		\includegraphics[width=0.45\linewidth]{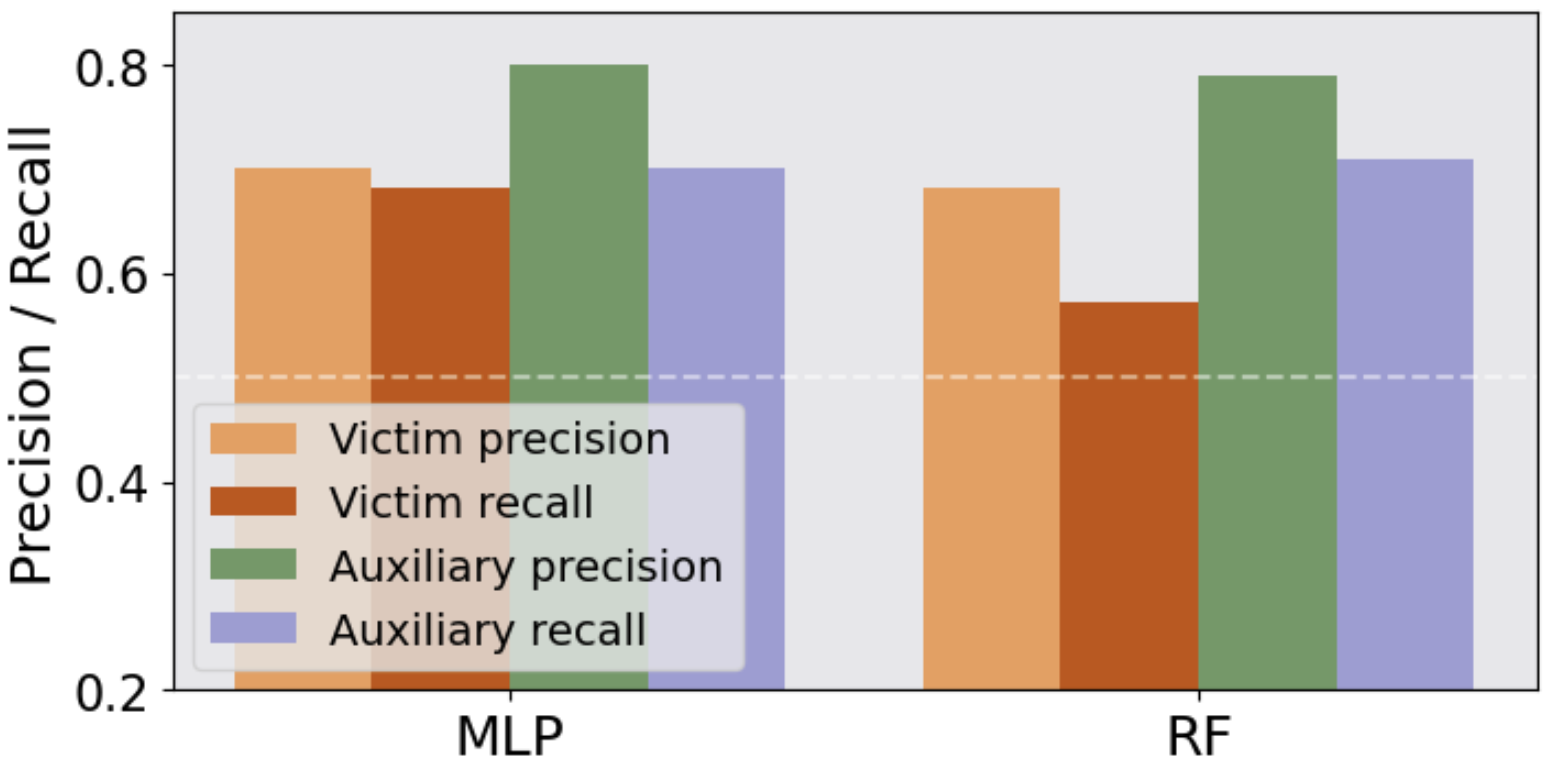}
		\label{fig:a}}
	\end{subfigure}
    \begin{subfigure}[Agnews]{
		\includegraphics[width=0.45\linewidth]{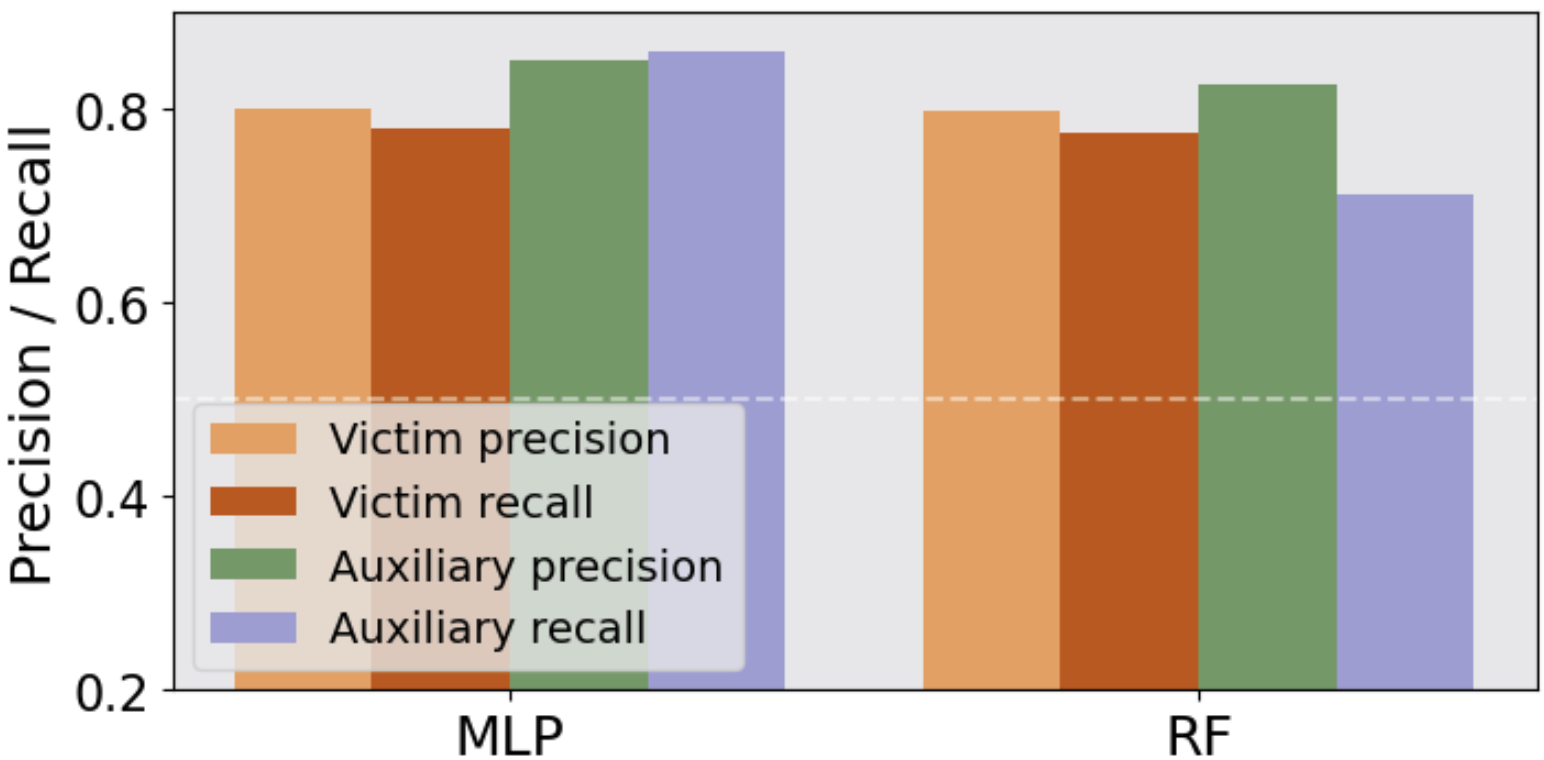}
		\label{fig:b}}
	\end{subfigure}
    \begin{subfigure}[20Newsgroup]{
		\includegraphics[width=0.45\linewidth]{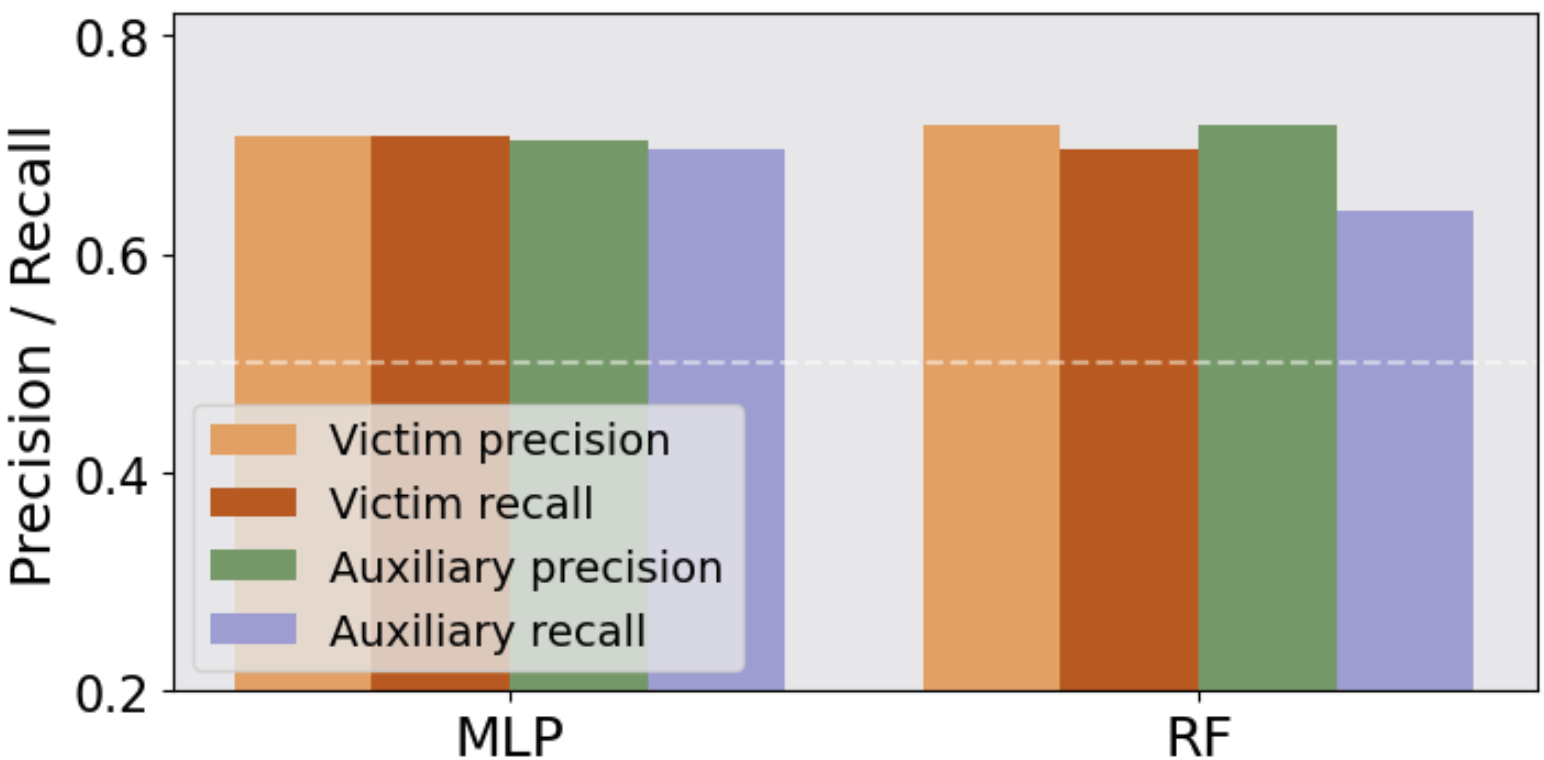}
		\label{fig:c}}
	\end{subfigure}
    \begin{subfigure}[Bank77]{
		\includegraphics[width=0.45\linewidth]{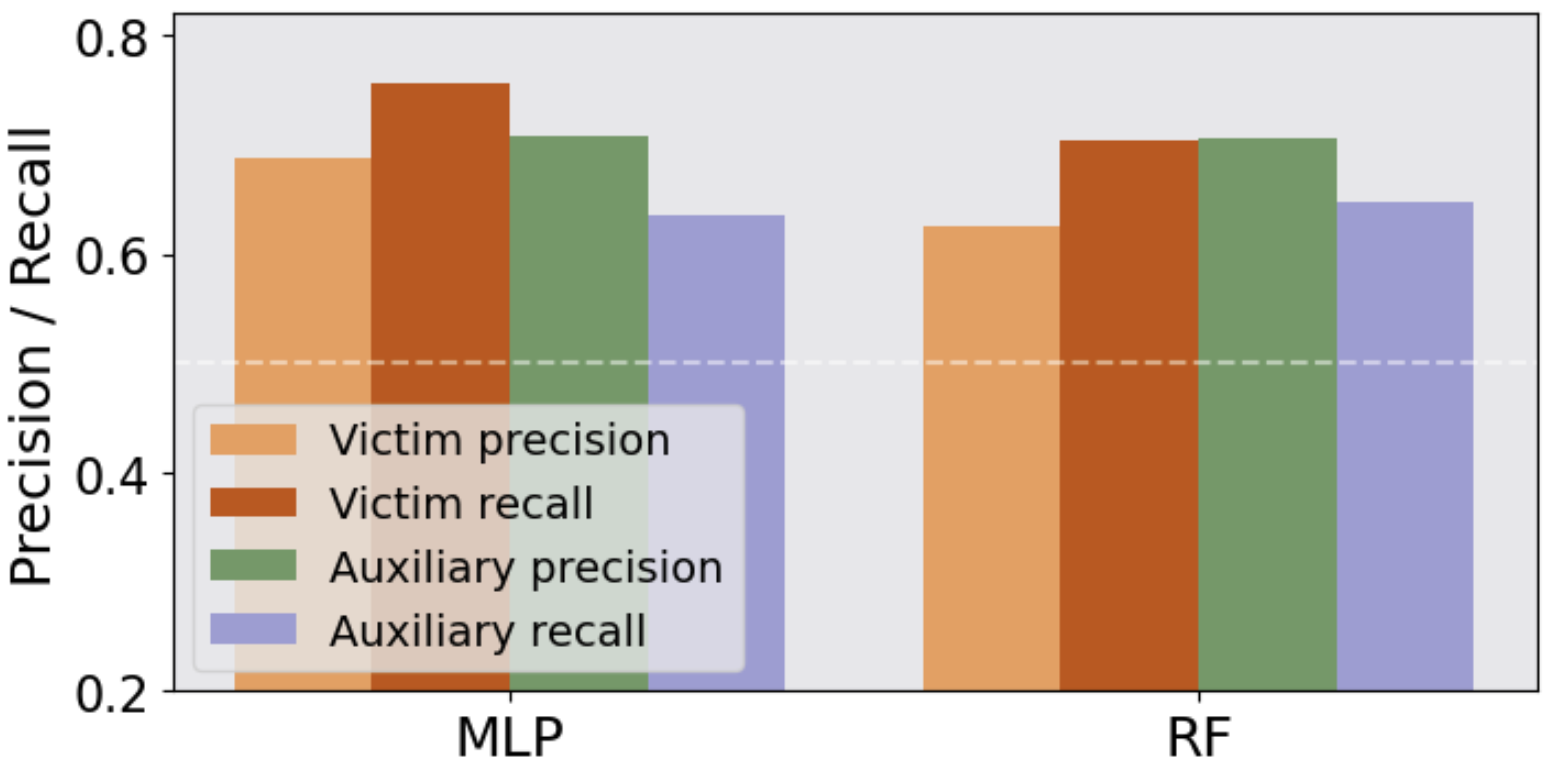}
		\label{fig:d}}
	\end{subfigure}
\caption{The MIA precision and recall on the victim and auxiliary datasets.}
\label{fig:attack3}
\end{figure}

\begin{figure*}[htbp]
\centering
	\begin{subfigure}[IMDB]{
		\includegraphics[width=0.24\linewidth]{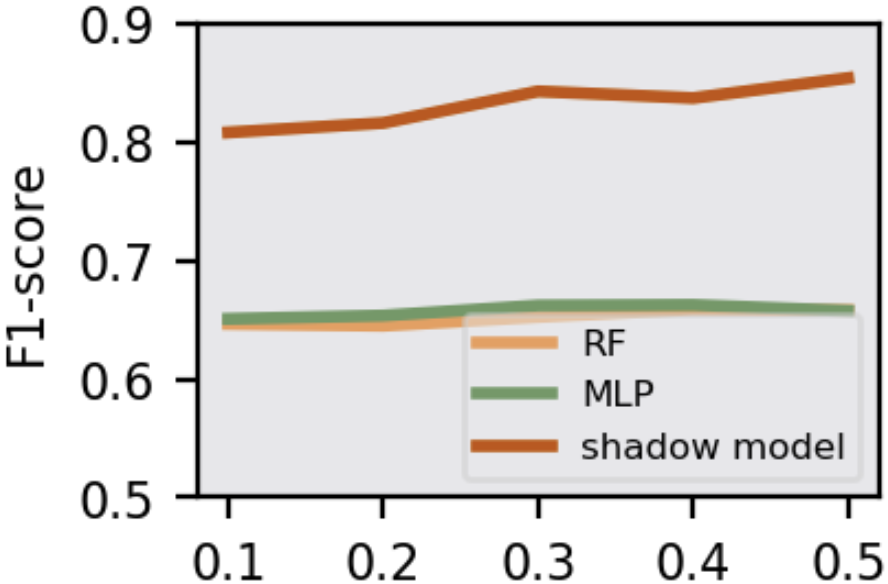}
		\label{fig:a}}
	\end{subfigure}
    \begin{subfigure}[Agnews]{
		\includegraphics[width=0.24\linewidth]{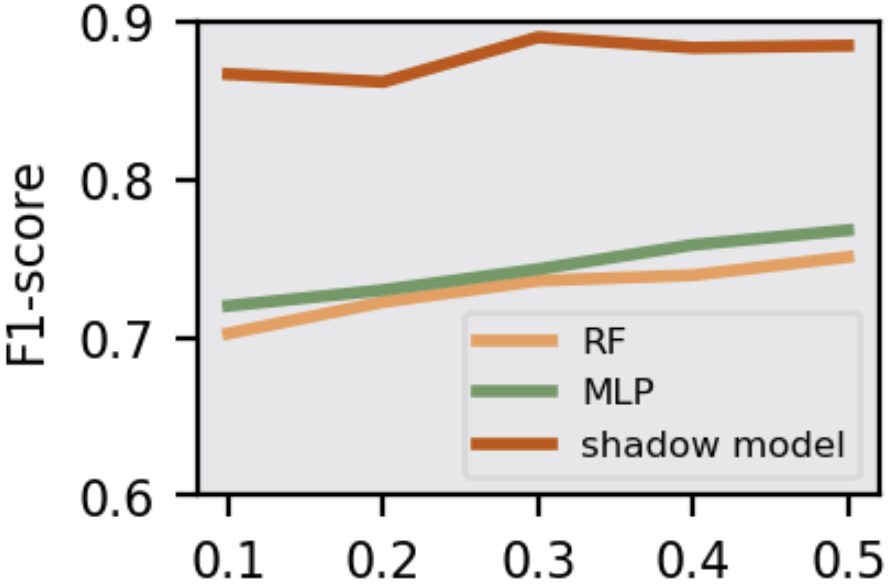}
		\label{fig:b}}
	\end{subfigure}
    \begin{subfigure}[20Newsgroup]{
		\includegraphics[width=0.24\linewidth]{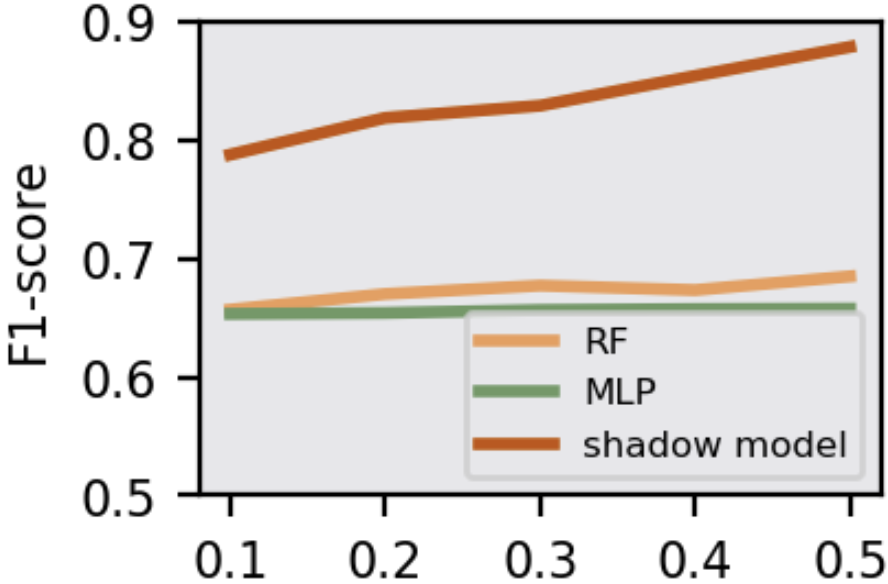}
		\label{fig:c}}
	\end{subfigure}
    \begin{subfigure}[Bank77]{
		\includegraphics[width=0.24\linewidth]{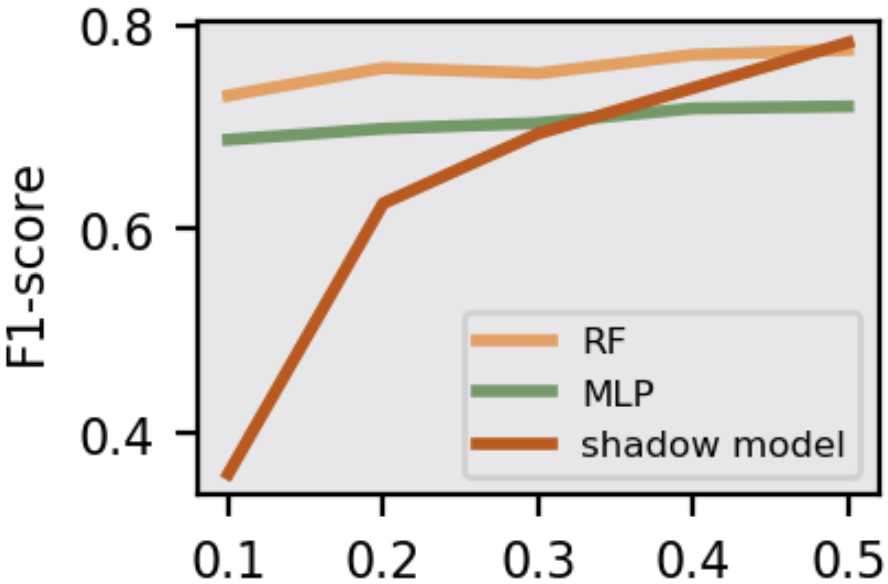}
		\label{fig:d}}
	\end{subfigure}
    \begin{subfigure}[WOS]{
		\includegraphics[width=0.24\linewidth]{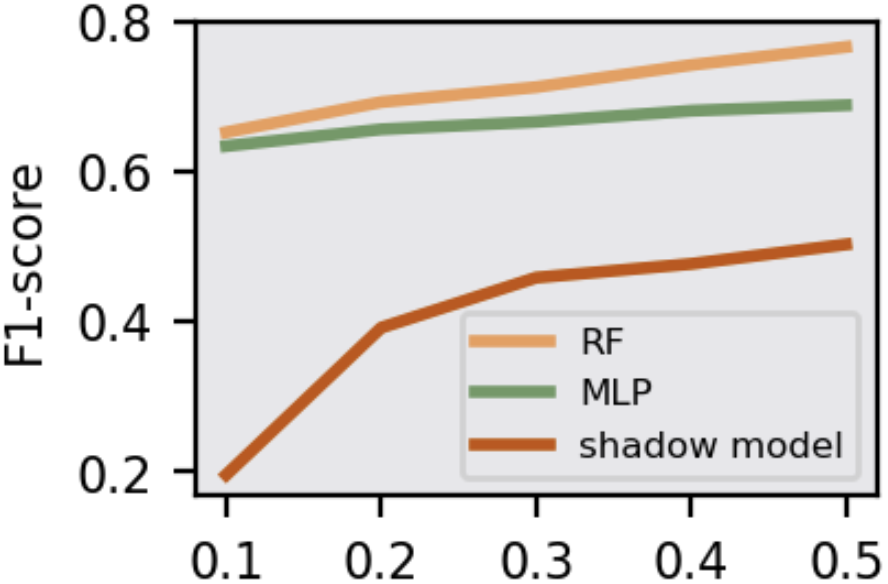}
		\label{fig:e}}
	\end{subfigure}
\caption{The F1-score of attack models and shadow models in Adversary 4.}
\label{fig:attack4}
\end{figure*}

\subsection{Classification Task}
Although \textit{PerProb} substantiates LLMs’ vulnerability to MIA in generation tasks, its reliance on indirect evaluation limits the impact quantification. To address this limitation and obtain a more tangible understanding of LLM memorization, we evaluate MIA through classification tasks using GPT-2, which provides direct ground truth labels for analysis. 

\textbf{Adversary 1 and Adversary 2.} Across five datasets, Adversaries 1 and 2 achieve average F1-scores of 71.41\% and 73.66\%, respectively, demonstrating successful MIA. While Adversary 1 performs well, it slightly lags behind Adversary 2, as the latter benefits from learned parameters of the $\mathcal{V}$. Interestingly, the performance of the attack model $\mathcal{A}$ does not improve as the number of categories in the datasets increases, suggesting that dataset complexity has minimal impact on MIA performance in classification tasks.

\textbf{Adversary 3.} The outcomes of Adversary 3 are depicted in Figure \ref{fig:attack3}, the precision of attack model exhibits a reduced around 5\% compared to Adversary 1. The results highlight that the performance of MIA on the $D_{victim}$ is closely linked to the characteristics of $D_{aux}$. In the Agnews dataset, $D_{aux}$ (20Newsgroup) shares thematic with the $D_{victim}$, the attack model achieves higher precision (79.86\%) and recall (77.63\%) on the $D_{victim}$, suggesting that the $D_{aux}$ effectively reinforces the $\mathcal{S}$'s ability to mimic the $\mathcal{V}$’s behavior. In contrast, $D_{aux}$ provides limited assistance for datasets with less similarity (e.g., IMDB and Bank77), leading to relatively lower MIA effectiveness. 

\textbf{Adversary 4.} The performance of the $\mathcal{S}$ (red line) and the attack models ($\mathcal{A}$), RF (yellow line), and MLP (green line) is shown in Figure \ref{fig:attack4}. As expected, the F1-score of the $\mathcal{S}$ improves consistently with the increase in training data from $D_{victim}$, and an upward trend in $\mathcal{A}$'s F1-score is observed. However, $\mathcal{S}$ performs worse in the Bank77 and WOS datasets due to dataset characteristics. Bank77’s average text length (56) makes achieving high performance challenging with limited data, while WOS’s over a hundred subcategories lead to confusion in $\mathcal{S}$ predictions. Interestingly, even when $\mathcal{S}$ performance is suboptimal, $\mathcal{A}$ achieves F1-scores above random chance, exceeding 60\% across all datasets and surpassing 70\% in the Agnews dataset. This suggests that the $\mathcal{S}$’s precision has a limited impact on $\mathcal{A}$’s effectiveness, as the posterior distributions generated by $\mathcal{S}$ still encode sufficient information for membership inference. 

Furthermore, RF and MLP exhibit slightly different trends across datasets. RF demonstrates robustness in datasets with more structured data, such as Bank77, while MLP shows advantages in datasets like Agnews and WOS, where the data distribution may involve non-linear complexities. This highlights the importance of selecting appropriate attack models based on dataset characteristics.

\begin{tcolorbox}[colback=gray!30, colframe=gray!70]
    \textbf{\textit{Finding 2}}: Auxiliary datasets with the same theme improve the effectiveness of the attack model. Moreover, attack models can successfully perform MIA even with minimal leakage of victim data.
\end{tcolorbox}

\section{Defenses} 
In this section, we evaluate common privacy-preserving techniques against MIA, including Knowledge Distillation (KD) \cite{gou2021knowledge}, Early Stopping (ES) \cite{dodge2020fine}, and Differential Privacy (DP) \cite{abadi2016deep}. KD and ES are applied in generative tasks, while DP is used in classification tasks, analyzing their effectiveness across all four attack strategies.

\subsection{Techniques}
\subsubsection{Knowledge Distillation.} 
KD is a widely used technique for transferring knowledge from a large teacher model to a smaller student model, which helps reduce overfitting and enhance generalization. In this study, the GPT-2 medium model is as the student model for GPT-2, and we choose GPT-Neo 2.7B as the teacher model, and GPT-Neo 1.3B is used as the student model for both $\mathcal{V}$ and $\mathcal{S}$. Therefore, the defense effect on the GPT-Neo 1.3B model is not discussed.  The distillation process uses “soft labels” produced by the teacher model, with a temperature of 2, to smooth the output distribution and ensure effective training \cite{li2023curriculum}. For generative tasks, the Kullback-Leibler divergence loss between the logits of the teacher and student models is minimized during training. Theoretically, the KD student model reduces dependency on pre-training datasets, potentially lowering memorization risks. To validate its generalization, we conduct cross-validation on different subsets of the pre-training datasets. 

\subsubsection{Early Stopping.} 
Early Stopping (ES) is a technique to prevent the model from overfitting the training data by halting training once performance stabilizes, thereby also preserving the model’s generalization ability. Theoretically, ES reduces the model’s reliance on specific training samples and mitigates memorization risks. To evaluate overfitting, we monitor the $PPL$ during training, which should decrease consistently. We set the ES threshold from 0.001 to 0.01 to find the optimal stopping point. If the decrement in $PPL$ falls below the threshold in consecutive training steps, we define this as the best ES point, signaling that further training may not improve generalization and could lead to overfitting.

\subsubsection{Differential Privacy.} 
Differential Privacy (DP) is a robust technique for safeguarding privacy by strategically injecting controlled noise into data, balancing data availability and confidentiality. While noise perturbation can degrade model performance, it mitigates MIA risks when prioritizes privacy over accuracy. In our implementation, noise is added to the output posteriors during the model training phase to minimize perturbation in the training process. We adopt the Laplace distribution as the noise distribution due to its ability to provide well-calibrated perturbations, defined as:

\begin{equation}
    f(x|\mu, b)=\frac{1}{2b}exp(-\frac{|x-\mu|}{b}),
\end{equation} 

where $\mu$ is the mean, $b$ is the scale parameter that controls the spread of the distribution, and $\epsilon$ controls the trade-off between privacy and data utility. $D \sim Lap(\mu, \frac{\Delta f}{\epsilon})$ is a Laplace distribution and satisfies the $\epsilon$-differential privacy. $\Delta f$ is the sensitivity, typically assigned a value of 1, which measures the maximum variation in the output of a function $f$ due to a unit change in its input. 

Traditionally, Laplace uses 0 as $\mu$, which minimally impacts posteriors in classification tasks and with few effects to defend MIA. To address this limitation, we introduce an adaptive mechanism using the maximum posterior value as the $\mu$ for noise generation. When the maximum posterior is high or low, the noise distribution is adjusted accordingly, ensuring appropriate perturbation to resist MIA. We evaluate traditional and improved DP across four attack patterns. The privacy budget $\epsilon$ is set to 0.5, 1, and 2 for evaluation, which represents a range of privacy levels, with lower values offering stronger privacy protection at the cost of more noise and higher values allowing for less noise and more accurate model outputs.

\begin{table*}[!tbp]
\centering
\caption{KD and ES efficacy in four attack patterns on GPT-2.}
\label{tab:de_gpt2}
\renewcommand{\arraystretch}{1.3}
\setlength{\tabcolsep}{1.5mm}{
\begin{tabular}{|c|c|c|c|c|c|c|c|c|c|c|c|c|}
    \toprule
    \multirow{3}{*}{\textbf{Dataset}} & \multicolumn{12}{c|}{\textbf{Adversaries / Defenses}}
    \\ \cline{2-13} 
    & \multicolumn{2}{c}{Adversary 1} & \multicolumn{2}{c}{\makecell{Knowledge \\ Distillation}} & \multicolumn{2}{c|}{Early Stopping} 
    & \multicolumn{2}{c}{Adversary 2} & \multicolumn{2}{c}{\makecell{Knowledge \\ Distillation}} & \multicolumn{2}{c|}{Early Stopping} 
    \\ \cline{1-13}
    & $PPL$ & $\lambda(W)$ & $PPL$ & $\lambda(W)$ & $PPL$ & $\lambda(W)$ & $PPL$ & $\lambda(W)$ & $PPL$ & $\lambda(W)$ & $PPL$ & $\lambda(W)$ 
    \\ \cline{2-13}
    \rowcolor{gray!40}
    IMDB    & 252.22 & -18.56 & 201.83 & \textbf{-23.5}  & \textbf{478.23} & -19.52  & 268 & -18.22 & 247.86 & \textbf{-23.51}  & \textbf{349.89} & -19.84        \\
    Agnews  & 1149.91 & -38.76 & 205.35 & \textbf{-51.32}  & \textbf{1483.73} & -35.84 & 915.69 & -33.26 & 926.18 & \textbf{-51.1}  & \textbf{1617.98} & -36.34   \\
    \rowcolor{gray!40} 
    WOS     & 240.17 & -24.62 & 179.23 & \textbf{-29.25}  & 245.27 & \textbf{-26.49}  & 162.59 & -25.13 & 176.79 & \textbf{-29.32}  & 206.03 & \textbf{-27.09}     \\
    \bottomrule
    \toprule
    \textbf{Dataset} & \multicolumn{2}{c}{Adversary 3} & \multicolumn{2}{c}{\makecell{Knowledge \\ Distillation}} & \multicolumn{2}{c|}{Early Stopping} 
    & \multicolumn{2}{c}{Adversary 4} & \multicolumn{2}{c}{\makecell{Knowledge \\ Distillation}} & \multicolumn{2}{c|}{Early Stopping}  
    \\ \cline{1-13}
    & $PPL$ & $\lambda(W)$ & $PPL$ & $\lambda(W)$ & $PPL$ & $\lambda(W)$ & $PPL$ & $\lambda(W)$ & $PPL$ & $\lambda(W)$ & $PPL$ & $\lambda(W)$ 
    \\ \cline{2-13}
    \rowcolor{gray!40}
    IMDB    & 307.05 & -19.67 & 221.22 & \textbf{-23.52}  & \textbf{381.83}  & -20.8  & 328.68 & -18.33 & 188.21 & \textbf{-23.54}  & \textbf{346.03} & -19.38        \\
    Agnews  & 1935.89 & -25.96 & 826.64 & \textbf{-50.4}  & 1078.96 & -33.19  & 1405.53 & -31.69 & 947.22 & \textbf{-50.92}  & \textbf{1676.84}  & -36.38   \\
    \rowcolor{gray!40} 
    WOS     & 269.39 & -23.2 & 198.5 & \textbf{-29.46}  & 263.93 & -26.18  & 220.07 & -24.96 & \textbf{364.98}  & \textbf{-29.35}  & 236.99  & -26.82     \\
    \bottomrule
\end{tabular}}
\end{table*}

\begin{table*}[!tbp]
\centering
\caption{KD and ES efficacy in four attack patterns on GPT-Neo 2.7B.}
\label{tab:de_gpt2.7}
\renewcommand{\arraystretch}{1.3}
\setlength{\tabcolsep}{1.75mm}{
\begin{tabular}{|c|c|c|c|c|c|c|c|c|c|c|c|c|}
    \toprule
    \multirow{3}{*}{\textbf{Dataset}} & \multicolumn{12}{c|}{\textbf{Adversaries / Defenses}}
    \\ \cline{2-13} 
    & \multicolumn{2}{c}{Adversary 1} & \multicolumn{2}{c}{\makecell{Knowledge \\ Distillation}} & \multicolumn{2}{c|}{Early Stopping} 
    & \multicolumn{2}{c}{Adversary 2} & \multicolumn{2}{c}{\makecell{Knowledge \\ Distillation}} & \multicolumn{2}{c|}{Early Stopping} 
    \\ \cline{1-13}
    & $PPL$ & $\lambda(W)$ & $PPL$ & $\lambda(W)$ & $PPL$ & $\lambda(W)$ & $PPL$ & $\lambda(W)$ & $PPL$ & $\lambda(W)$ & $PPL$ & $\lambda(W)$ 
    \\ \cline{2-13}
    \rowcolor{gray!40}
    IMDB    & 1.08 & -7.11 & 1.27 & \textbf{-11.46} & \textbf{3.11} & -10.77 & 1.61 & -5.94 & 1.13 & \textbf{-9.42} & \textbf{3.98} & -6.02   \\
    Agnews  & 9.96 & -10.15 & \textbf{184.08} & -9.68 & 12.30 & \textbf{-13.28} & 8.49 & -10.03 & \textbf{266.87} & -10.20 & 9.01 & \textbf{-10.22}   \\
    \rowcolor{gray!40} 
    WOS     & 110.23 & -10.02 & 57.92 & \textbf{-16.51} & \textbf{102.03} & -11.79 & 1.16 & -6.32 & 1.15 & -6.40 & \textbf{3.32} & \textbf{-11.68}   \\
    \bottomrule
    \toprule
    \textbf{Dataset} & \multicolumn{2}{c}{Adversary 3} & \multicolumn{2}{c}{\makecell{Knowledge \\ Distillation}} & \multicolumn{2}{c|}{Early Stopping} 
    & \multicolumn{2}{c}{Adversary 4} & \multicolumn{2}{c}{\makecell{Knowledge \\ Distillation}} & \multicolumn{2}{c|}{Early Stopping}  
    \\ \cline{1-13}
    & $PPL$ & $\lambda(W)$ & $PPL$ & $\lambda(W)$ & $PPL$ & $\lambda(W)$ & $PPL$ & $\lambda(W)$ & $PPL$ & $\lambda(W)$ & $PPL$ & $\lambda(W)$ 
    \\ \cline{2-13}
    \rowcolor{gray!40}
    IMDB    & 8.16 & -10.65 & 3.26 & -7.89 & \textbf{8.07} & \textbf{-10.13} & 8.08 & -8.01 & 10.08 & \textbf{-14.70} & \textbf{11.24} & -10.53   \\
    Agnews  & 21.32 & -11.97 & 1.88 & -11.38 & 33.01 & \textbf{-19.27} & 10.96 & -10.05 & 51.69 & \textbf{-19.25} & \textbf{156.52} & -10.48   \\
    \rowcolor{gray!40} 
    WOS     & 248.02 & -8.66 & 203.50 & -14.12 & \textbf{322.75} & \textbf{-14.36} & 24.10 & -7.37 & 16.69 & -7.83 & \textbf{23.72} & \textbf{-9.97}   \\
    \bottomrule
\end{tabular}}
\end{table*}

\subsection{Result} 
\subsubsection{Generation Problem} 
The results in Table \ref{tab:de_gpt2} illustrate KD and ES effectively mitigate MIA susceptibility for \textbf{GPT-2}. Generally, both defense mechanisms result in higher $PPL$ and lower $\lambda(W)$ for adversarially generated datasets across all attack patterns. In Adversary 1 and Adversary 2, for both KD and ES, the increase in $PPL$ and the decrease in $\lambda(W)$ are evident, reflecting reduced alignment between the $\mathcal{V}$ and generated datasets. Especially, the $PPL$ of Agnews increased to 1483.73 in Adversary 1 under ES strategy, as ES disrupts adaptation to specific patterns in training data.

However, in Adversary 3, the $PPL$ exhibits a decrease in KD, particularly in the Agnews dataset, dropping from 1935.89 to 826.64. This may be due to KD enhancing alignment by leveraging $D_{aux}$ similarities and improving the $\mathcal{S}$'s generalization capability. Specifically, the $D_{aux}$ used during training may semantically overlap with the $D_{victim}$, as seen in Agnews, where thematic similarities exist between $D_{aux}$ (20Newsgroup) and $D_{victim}$. KD allows the $\mathcal{S}$ to leverage these similarities, generating more general data that the $\mathcal{V}$ interprets with lower $PPL$ compared to the original attack pattern. However, when combining $\lambda(W)$, the KD effect may not necessarily be counterproductive, as the decrease in $\lambda(W)$ indicates strong resistance to MIA. Consequently, the result underscores a nuanced limitation of KD that while it effectively reduces overfitting and enhances generalization, it may inadvertently improve alignment between $\mathcal{V}$ and $\mathcal{S}$ in Adversary 3. Therefore, the choice of defense mechanism according to different attack patterns could affect effectiveness. Although the $\lambda(W)$ still shows resistance on MIA, the choice of $D_{aux}$ should still avoid the same theme.

Both KD and ES achieve similar effectiveness in increasing $PPL$ and reducing $\lambda(W)$, indicating resistance to MIA even when partial victim data is used in $\mathcal{S}$. Besides, except for Adversary 4, both Adversary 1 and 2 exhibited that KD outperforms ES in reducing $\lambda(W)$, while ES achieves higher $PPL$, reflecting their differing mechanisms. KD primarily focuses on transferring generalized knowledge from the teacher to the student model, which helps reduce overfitting to specific data points and lowers the $\mathcal{V}$’s confidence when evaluating generated data. In contrast, ES prevents the $\mathcal{V}$ from overfitting during training by halting the training process early, disrupting the model’s adaptation to the finer details of the training set, leading to an overall increase in $PPL$. 

Table \ref{tab:de_gpt2.7} shows a similar trend of KD and ES improving MIA resistance for \textbf{GPT-Neo 2.7B}. However, compared to GPT-2, the results on GPT-Neo 2.7B show that KD is consistently superior across most datasets and attack scenarios. The increased capacity of GPT-Neo 2.7B likely enables more effective distillation and generalization, and enhances resistance to MIA, as reflected in higher $PPL$ and lower $\lambda(W)$. In Adversary 1 with Agnews, $PPL$ increases to 184.08, and $\lambda(W)$ from -10.15 slightly rises to -9.68. This minor increase in $\lambda(W)$ does not necessarily indicate ineffective defenses but rather reflects the interplay between the $\mathcal{V}$'s generalization capability and the semantic alignment of the $\mathcal{S}$’s generated data. However, this also highlights a potential vulnerability that adversaries can exploit this alignment to slightly enhance the attack’s success, particularly when the dataset is semantically simple and well-aligned. The same situation is also observed in Adversary 3, on the IMDB dataset, from -10.65 rise to -7.89. Meanwhile, the downward $PPL$ is observed, like the findings in GPT-2, the $PPL$ decline from 21.32 to 1.88 in the Agnews dataset, further proving the importance of select defenses for distinct attack patterns. In Adversary 4, KD and ES both provide a robust defense, though KD slightly outperforms ES in $\lambda(W)$, and ES outperforms in $PPL$, where in the IMDB dataset, KD achieves $\lambda(W)$ of -23.54 and $PPL$ of 346.03.

When comparing the two models, the results highlight that GPT-Neo 2.7B generally exhibits stronger resistance to MIA due to its capacity to generalize better and avoid overfitting. KD amplifies this advantage, as its ability to smooth output distributions aligns well with the model’s intrinsic strengths. Besides, GPT-Neo 2.7B benefits more from ES in scenarios with simpler data distributions, where overfitting is a more immediate concern. The dataset characteristics also play a crucial role, as thematically consistent datasets in Adversary 3 for KD may bring improvement in $PPL$. In general, the findings underscore the interaction between model architecture, dataset complexity, and defense strategies, emphasizing the need for customized approaches to effectively mitigate MIA risks.

\begin{tcolorbox}[colback=gray!30, colframe=gray!70]
    \textbf{\textit{Finding 3}}: Both KD and ES effectively mitigate MIA risks in LLMs, with ES showing superior performance in elevating $PPL$, while KD performs better in lowering $\lambda(W)$. Besides, highlights the importance of selecting appropriate defense strategies against different attack patterns.
\end{tcolorbox}

\subsubsection{Classification Problem}

\begin{table*}[!tbp]
\centering
\caption{Traditional and improved DP efficacy in all datasets. The best performance settings are highlighted.}
\label{tab:de_dp}
\renewcommand{\arraystretch}{1.6}
\setlength{\tabcolsep}{1.2mm}{
\begin{tabular}{|cc|c|c|c|c|c|c|c|c|c|c|c|c|c|c|}
    \toprule
    \multicolumn{2}{|c|}{\multirow{2}{*}{\textbf{DP}}}  & \multicolumn{14}{c|}{\textbf{Datasets}} \\
    \cline{3-16}
     & & \multicolumn{3}{c|}{\textbf{\textit{IMDB}}} & \multicolumn{3}{c|}{\textbf{\textit{Agnews}}} & \multicolumn{3}{c|}{\textbf{\textit{20Newsgroup}}} & \multicolumn{3}{c|}{\textbf{\textit{Bank77}}} & \multicolumn{2}{c|}{\textbf{\textit{WOS}}} \\
    
     $\mu$ & $\epsilon$ & Adv.1 & Adv.2 & Adv.3 & Adv.1 & Adv.2 & Adv.3 & Adv.1 & Adv.2 & Adv.3 
       & Adv.1 & Adv.2 & Adv.3 & Adv.1 & Adv.2 \\
    \hline
    0   &  0.5  & 0.644 & 0.637 & 0.567 & \textbf{0.706} & 0.723 & 0.708 & 0.604 & 0.602 & 0.609 & 0.725 & 0.730 & 0.676 & 0.549 & 0.554 \\
    \rowcolor{gray!40}
    0   &   1   & 0.639 & 0.640 & 0.582 & 0.735 & 0.729 & 0.693 & 0.606 & 0.608 & \textbf{0.600} & 0.728 & 0.736 & 0.678 & 0.545 & 0.560 \\
    0   &   2   & 0.642 & 0.644 & 0.581 & 0.741 & 0.746 & 0.680 & 0.614 & 0.613 & 0.621 & 0.733 & 0.734 & 0.688 & 0.560 & 0.573 \\
    \hline
    \rowcolor{gray!40}
    $M^{\mathrm{a}}$ &  0.5  & \textbf{0.626} & 0.638 & \textbf{0.520} & 0.714 & \textbf{0.716} & 0.633 & \textbf{0.572} & \textbf{0.583} & 0.616 & 0.722 & \textbf{0.717} & \textbf{0.640} & \textbf{0.525} & \textbf{0.552} \\
    $M^{\mathrm{a}}$ &   1   & 0.635 & 0.636 & 0.529 & 0.730 & 0.737 & \textbf{0.586} & 0.580 & 0.585 & 0.648 & \textbf{0.721} & 0.724 & 0.678 & 0.547 & 0.572 \\
    \rowcolor{gray!40}
    $M^{\mathrm{a}}$ &   2   & 0.634 & \textbf{0.633} & 0.534 & 0.742 & 0.751 & 0.690 & 0.597 & 0.599 & 0.685 & 0.724 & 0.727 & 0.677 & 0.556 & 0.566 \\
    \hline   
    \multicolumn{2}{|c|}{Original} &
                  0.652 & 0.686 & 0.602 & 0.765 & 0.800 & 0.789 & 0.711 & 0.710 & 0.707 & 0.753 & 0.776 & 0.691 & 0.689 & 0.711 \\
    \bottomrule
    \multicolumn{16}{l}{$^{\mathrm{\textit{a}}}$ The max value of posteriors.}  
\end{tabular}}
\end{table*}

The outcomes are summarized in Table \ref{tab:de_dp}. Our findings demonstrate that DP's effectiveness varies across datasets and adversarial scenarios, especially the improved DP. In the first three adversary patterns, DP performs better on the 20Newsgroup and WOS datasets than others, which improve by 12.43\% and 16.15\%, respectively, likely due to their unique characteristics. For 20Newsgroup, it's moderate text length and balanced class distribution make posterior probabilities more susceptible to DP-induced noise. In WOS, many subcategories amplify the confusion introduced by noise, resulting in stronger defenses. In Adversary 2, DP achieves a modest F1-score reduction of approximately 9.64\% due to the adversaries' access to the $\mathcal{V}$'s internal parameters, which limits DP's impact. In Adversary 3, DP demonstrates significant effectiveness on the Agnews dataset (accuracy reduction of 20\%) when the $D_{aux}$ (20Newsgroup) shares a similar theme. However, DP achieves less than 10\% accuracy reduction for datasets with greater distributional differences, indicating limited effectiveness. These findings suggest that DP is particularly effective in black-box attacks where $D_{aux}$ are similar to $D_{victim}$ or when datasets have structural features conducive to noise-induced confusion. Moreover, the improved DP method demonstrates enhanced defense capabilities, particularly in datasets with higher category complexity, and lower privacy budgets (smaller $\epsilon$) further amplify DP's effectiveness, albeit with potential trade-offs in model performance.

\begin{figure*}[htbp]
\centering
	\begin{subfigure}[IMDB]{
		\includegraphics[width=0.24\linewidth]{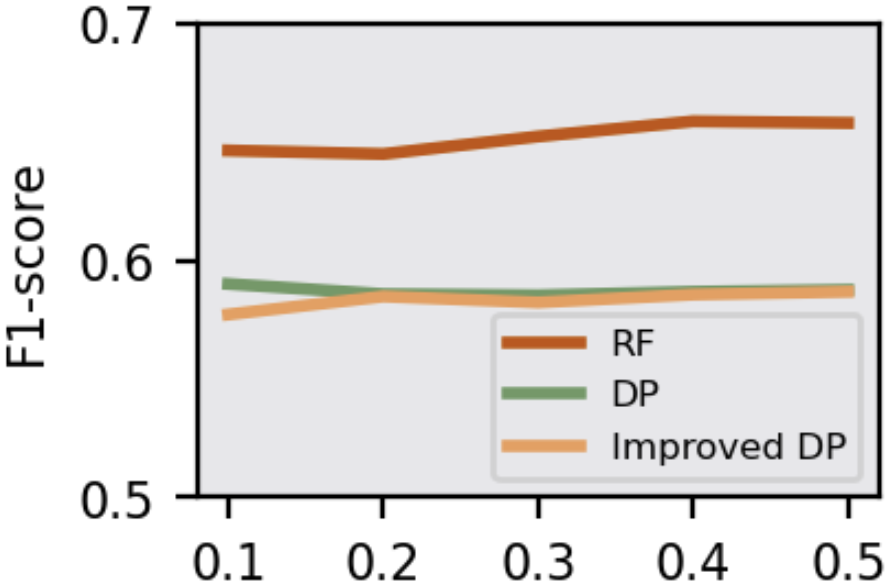}
		\label{fig:a}}
	\end{subfigure}
    \begin{subfigure}[Agnews]{
		\includegraphics[width=0.24\linewidth]{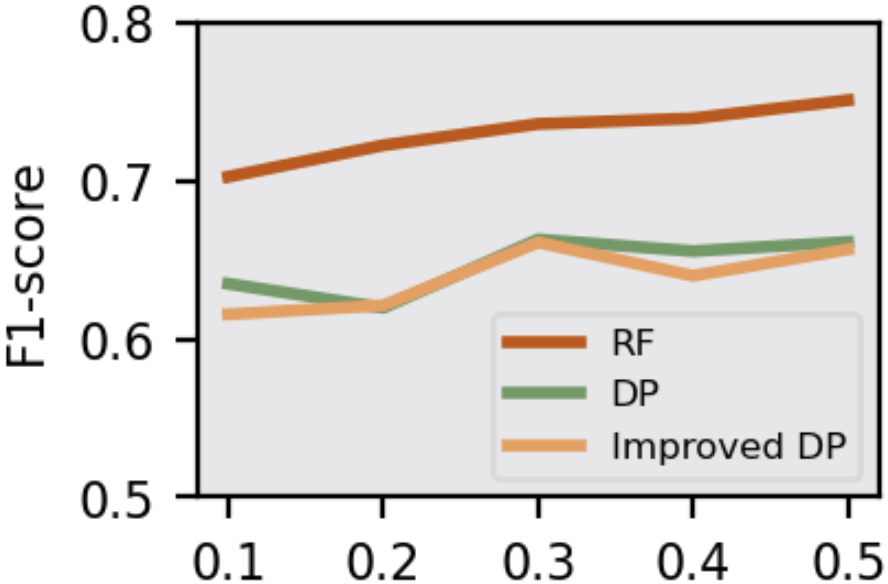}
		\label{fig:b}}
	\end{subfigure}
    \begin{subfigure}[20Newsgroup]{
		\includegraphics[width=0.24\linewidth]{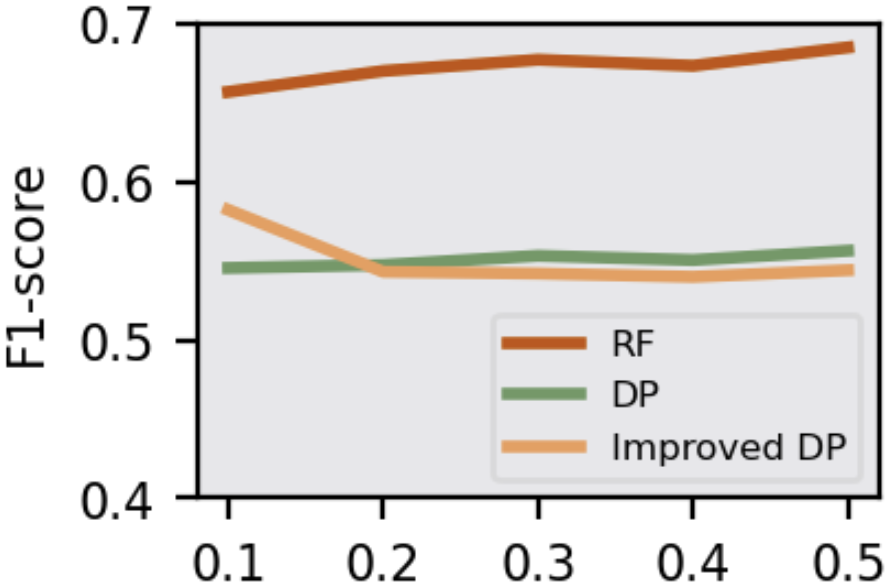}
		\label{fig:c}}
	\end{subfigure}
    \begin{subfigure}[Bank77]{
		\includegraphics[width=0.24\linewidth]{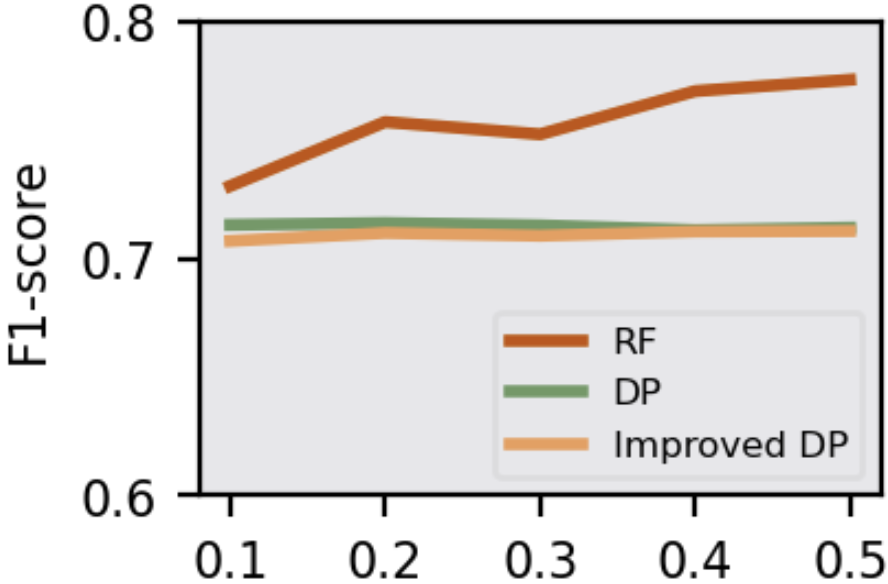}
		\label{fig:d}}
	\end{subfigure}
    \begin{subfigure}[WOS]{
		\includegraphics[width=0.24\linewidth]{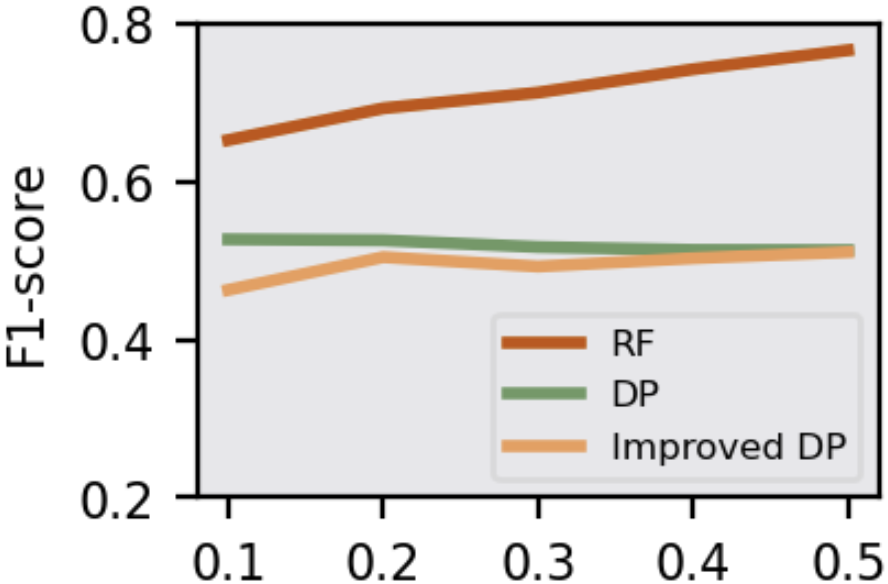}
		\label{fig:e}}
	\end{subfigure}
\caption{F1-score of attack models in Adversary 4 under DP mechanism.}
\label{fig:defense}
\end{figure*}

The F1-scores of Adversary 4 under the DP mechanism, as shown in Figure \ref{fig:defense}, highlight DP's effectiveness even in challenging white-box scenarios where partial $D_{victim}$ is used to train $\mathcal{S}$. Both traditional and improved DP mechanisms at lower thresholds induce significant errors in the $\mathcal{A}$, reducing F1-scores by up to 10\% in most datasets. Improved DP consistently slightly outperforms traditional DP, particularly in datasets with high category complexity like 20Newsgroup and WOS, where adaptive noise better obscures posterior distributions, while the Bank77 is excepted as the short length. However, as thresholds increase, the effectiveness of DP diminishes, with attackers gaining more insights from the increased victim data, particularly in datasets like Agnews, where the F1-score of $\mathcal{A}$ increases around 5\%. These findings suggest that DP, especially the improved DP, is most effective in scenarios with limited victim data or when applied to datasets with higher structural complexity.

\begin{tcolorbox}[colback=gray!30, colframe=gray!70]
    \textbf{\textit{Finding 4}}: Except for Adversary 3, DP can significantly defend MIA, and the efficacy of the improved DP method is more pronounced, which sets the max posteriors as the mean of the noise distribution.
\end{tcolorbox}

\section{Limitations}

While \textit{PerProb} provides a generalizable and label-free framework for evaluating LLM memorization, several limitations warrant further discussion. First, our experimental evaluation focuses on mid-sized open-source models such as GPT-2 and GPT-Neo, primarily due to resource constraints. Although these models are widely used in prior MIA studies, they may not fully reflect the memorization behavior of larger proprietary LLMs such as GPT-4 or Claude. Nonetheless, the core design of \textit{PerProb} that based on output-level $PPL$ and $\lambda(W)$ remains model-independent and can be applied to API-only or black-box LLMs, where internal access and member labels are unavailable. We leave the evaluation on larger-scale or closed-source LLMs as future work.

Although prior works such as shadow model attacks \cite{shokri2017membership} and confidence-based thresholding \cite{yeom2018privacy} have demonstrated success in conventional ML settings, they rely on access to labeled member/non-member data or internal model outputs. These requirements do not align with the constraints of real-world LLM evaluation, where neither the training data nor the internal confidence scores are typically available. As such, we do not directly compare against these baselines in our experiments, but summarize their limitations and distinctions in Table~\ref{tab:bg}.

Second, \textit{PerProb} currently assumes consistent prompts between shadow and victim models for generation tasks. While this assumption holds in controlled experimental designs, prompt variance in real-world deployments may affect metric stability. Incorporating prompt perturbation robustness or ensemble-based inference could improve reliability.

Third, although our classification experiments demonstrate measurable attack success, the evaluation focuses on specific datasets and attack models (RF and MLP). Future extensions can consider other classification tasks (e.g., summarization, code generation) and attacker models (e.g., GBDT, neural attackers) to further validate generality.

\section{Threats to Validity}

\textbf{Internal Validity.}  
Potential threats to internal validity arise from experimental control and data quality. First, dataset bias and text length may affect the reliability of memorization measurement. To mitigate this, we adopt five datasets with varying domains and sequence lengths to ensure a more representative evaluation. Second, ES is sensitive to the choice of validation data, which could result in premature convergence. We address this by validating across multiple datasets to avoid overfitting to any single distribution. Third, KD may lead to loss of generalization due to compression. To ensure robustness, we cross-validate KD results on multiple sub-datasets.

\textbf{External Validity.}  
Our experiments are conducted on GPT-2 and GPT-Neo (1.3B and 2.7B), which represent small to mid-scale LLMs. While these models are commonly used in prior MIA research, results may not fully generalize to larger proprietary models such as GPT-4 or Claude. However, the design of \textit{PerProb} is model-independent and can be applied to any LLM with output access. Future work will evaluate our approach on more diverse architectures.

\textbf{Construct Validity.}  
Construct validity concerns arise from whether the experimental setup truly reflects the model memorization. First, our task design may introduce performance bias, as LLM memorization varies between generation and classification settings. To reduce this, we explicitly separate the two task types in evaluation. Second, our use of RF and MLP as attack models may limit attack diversity. However, they are chosen due to their complementary strengths: RF handles large-scale tabular features and outliers effectively, while MLP is capable of modeling non-linear relationships and generalizing across scenarios.

\textbf{Conclusion Validity.}  
The stochasticity of DP may affect performance reproducibility. To balance privacy and accuracy, we tune the privacy budget to control degradation and ensure stability. All results are averaged over multiple runs to reduce variance and improve reliability. Overall, while our conclusions are derived from the widely-used settings, they remain subject to verification on more diverse model families and under broader attack scenarios in future work.

\section{Conclusion}
Due to ongoing debates about whether LLMs are susceptible to MIA, we proposed \textit{PerProb}, a novel framework for indirectly evaluating LLMs memorization by comparing $PPL$ and $\lambda(W)$ of datasets generated by shadow models that tested on a victim model. \textit{PerProb} enables systematic assessment of LLM vulnerability without requiring ground truth member/non-member data, making it a practical and scalable approach to understanding privacy risks.

Our findings offer key insights across generation and classification tasks. In generation tasks, larger models like GPT-Neo 2.7B exhibit more substantial generalization capabilities and inherent resistance to memorization, while smaller models like GPT-2 show higher susceptibility to overfitting and direct memorization of training data. In classification tasks, \textit{PerProb} revealed measurable memory traces in LLMs, with attack models consistently outperforming random guessing, underscoring the persistent privacy risks posed by MIA even in controlled scenarios. We also assessed defense mechanisms, including Knowledge Distillation (KD), Early Stopping (ES), and Differential Privacy (DP), all of which effectively increased $PPL$ and reduced $\lambda(W)$. KD excelled in lowering $\lambda(W)$ by transferring generalized knowledge and reducing model confidence, while ES more effectively elevated $PPL$ by mitigating overfitting. DP and its improved setting can both enhance MIA resistance, particularly in black-box settings or datasets with high category complexity, although their effectiveness varies with dataset characteristics and privacy budget settings. However, KD showed nuanced limitations, such as inadvertently alignment between shadow and victim models, thus improving the $PPL$ when semantically similar auxiliary datasets are used in Adversary 3. These findings emphasize the importance of defense strategy selection in different attack patterns and highlight the complementary strengths of KD, ES, and DP, suggesting their combined potential to mitigate MIA.

This study advances the understanding of MIA against LLMs by introducing \textit{PerProb} as a robust evaluation tool and offering insights into the interplay between model size, attack patterns, and defenses. \textit{PerProb} provides a novel, indirect approach to evaluating LLM memorization, offering a practical framework for assessing privacy risks. We will extend \textit{PerProb} to larger and more complex LLMs, explore advanced defenses tailored to LLMs, and examine the impact of dataset diversity on MIA risks in future work.

\bibliographystyle{IEEEtran}  
\bibliography{ref}

\end{document}